\documentclass[twocolumn,showpacs,preprintnumbers,amsmath,amssymb]{revtex4}
\usepackage{graphicx}% Include figure files
\usepackage{dcolumn}% Align table columns on decimal point
\usepackage{bm}% bold math
\def\bea {\begin{eqnarray}}
\def\eea {\end{eqnarray}}
\def\be {\begin{equation}}
\def\ee {\end{equation}}
\def\nn {\nonumber}
\begin{document}

\title{$\phi$ production at RHIC: characterization of co-existence phase}

\author{Jajati K. Nayak$^1$, Jan-e Alam$^1$, Bedangadas Mohanty$^1$,
Pradip Roy$^2$ and Abhee K. Dutt-Mazumder$^2$}

\medskip

\affiliation{(1) Variable Energy Cyclotron Centre, 1/AF, Bidhan Nagar , 
Kolkata - 700064}

\affiliation{(2) Saha Institute of Nuclear Physics, 1/AF, Bidhan Nagar , 
Kolkata - 700064}

\begin{abstract}

We  extract the effective degrees of freedom 
that characterize the co-existing phase of quark gluon plasma and 
hadrons. Experimental data on $\phi$ at 
mid-rapidity is used to set a  lower bound to the critical temperature
of quark hadron phase transition. The production and evolution 
of strangeness have been studied by using Boltzmann equation.
The results have been contrasted with the 
experimental data obtained by STAR collaboration at RHIC
for Au + Au collisions at $\sqrt{s_{NN}}=200$ GeV.
Our study reveals that the $\phi$ mesons freeze out 
at a temperature $\sim 160$ MeV, a value close to the 
transition  temperature for quark-hadron phase transition.

\end{abstract}

\pacs{25.75.-q,25.75.Dw,24.85.+p}

\maketitle

The goal of the ultra-relativistic heavy ion collisions 
is to create and study quark gluon plasma (QGP). 
The pre-requisite
for the formation of QGP is to create a hot and dense system 
of nuclear matter.
Experimental results from the Relativistic Heavy-Ion Collider (RHIC) 
have indicated the formation of such a state of  matter in the 
early stage~\cite{rhicwhitepapers}. Among others,
one of the promising signals of QGP formation is the enhanced production of
strange and anti-strange particles~\cite{rafelski}. As a result, the production of 
particles with hidden ({\it e.g.} $\phi$) as well as open strangeness have
become a field of intense theoretical~\cite{chen,pal,fuchs,letessier}
and experimental~\cite{phenix,STAR,STAR1,na49} activities.

In the present work we  study
the production of $\phi$ meson at RHIC. The $\phi$ meson is composed of a 
strange ($s$) and anti-strange ($\bar{s}$) quark and its interaction 
with nuclear matter is suppressed 
according to  Okubo-Zweig-Izuka (OZI) rule. In  QGP 
$s$ and $\bar{s}$ quarks are produced mainly by 
gluon fusion and annihilation of light ($u$ and $d$) quarks and anti-quarks.
These $s\bar{s}$ will form $\phi$ through the hadronization process.
This process
is not OZI suppressed. Therefore, we expect excess $\phi$ if
QGP is formed in the initial state of heavy ion collisions as 
compared to hadronic initial state. 
The production of $\phi$ mesons from sources
other than plasma is expected to be small, so they will not overshadow the
plasma signal. 
The other advantage of $\phi$
is that after its production during the hadronization of the QGP,  
it suffers less re-scattering in the hadronic matter, thereby retaining
information of the thermodynamic state of the matter during its
hadronization stage. Therefore,  study of $\phi$ spectra will be 
very useful to determine the transition temperature ($T_c$) for quark-hadron
phase transition.
However, it may be mentioned 
that the kaons which come from decay of $\phi$ mesons will 
undergo re-scattering, and the $\phi$'s experimentally reconstructed from
these kaons may carry some re-scattering effect. 
Still reliable information on 
$T_{c}$ can be extracted by measuring $\phi$ productions
in high energy heavy ion collisions~\cite{shor}.

Hence we shall show  the usefulness of $\phi$ as
a probe for co-existence phase by studying its production at 
$\sqrt{s_{NN}}$ = 200 GeV~\cite{STAR}. 
First we demonstrate from the experimental data that  $\phi$ 
meson freezes out at a temperature close to $T_{c}$, the value of which is
obtained  from  lattice QCD simulations~\cite{lattice}. 
We argue, using  the yield of $\phi$ mesons at mid-rapidity 
($\frac{dN_{\phi}}{dy}$) 
that the enhanced production is possible if 
the system has passed through a mixed phase of quarks, gluons and hadrons
and the temperature during the hadronization is $> 160$ MeV. 
Finally, attempt will be made to characterize the mixed phase by putting 
bounds on the the effective statistical degeneracy.

\begin{figure}
\begin{center}
\includegraphics[scale=0.5]{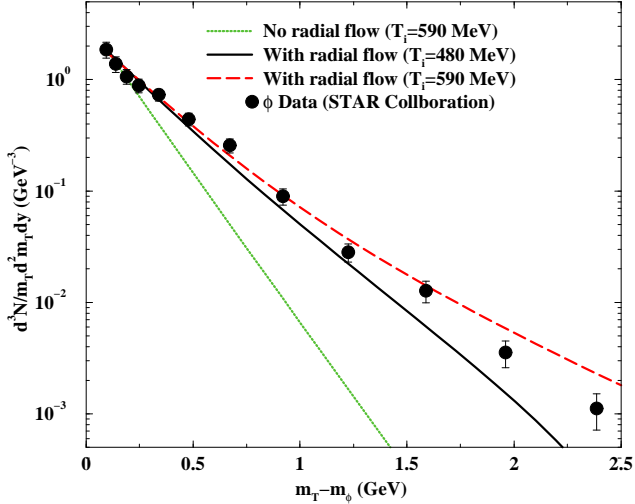}
\caption{
Transverse mass distribution of $\phi$ mesons for 0-5\% central
Au+Au collisions at $\sqrt{s_{NN}}$ = 200 GeV. Compared to hydrodynamical 
calculations with and without radial flow and for two different initial 
conditions. The equation of sate used is obtained from
lattice QCD results.
}
\label{fig1}
\end{center}
\end{figure}

Relativistic hydrodynamics with (3+1) dimensional expansion~\cite{hvg} 
and  boost invariance along the longitudinal direction~\cite{bjorken} 
has been used to study
the evolution of the matter formed in heavy ion collisions. 
We take the equation
of state from lattice QCD simulations~\cite{lattice}. The initial
velocity and energy density profiles are taken as $v(r,\tau_i)=0$ and
$\epsilon(r,\tau_i)=\epsilon_0/(1+exp(r-R)/\delta)$ respectively. Here
$\tau_i$ is the thermalization time, $R$ is nuclear radius. Results will
be shown for two sets of values of $\tau_i=0.15$ and $0.2$ fm/c. 
Such values of $\tau_i$~\cite{jane,peress} have recently been used 
to reproduce the thermal photon spectra measured by PHENIX collaboration
~\cite{phenixphoton} in Au + Au collisions at $\sqrt{s_{NN}}=200$ GeV.
The corresponding
values for the initial temperatures are 590 and  480 MeV respectively.
Results of transverse momentum spectra of $\phi$ mesons are shown in 
fig.~\ref{fig1}. 
It appears that the results for
$T_i=590$ MeV describes the data quite well for the entire range of 
transverse momentum ($p_T$) under consideration. However,
for $p_T$ more than 2 GeV (or $m_T-m_\phi\sim 1.2$ GeV, where
$m_T=\sqrt{p_T^2+m_\phi^2}$, is the transverse mass and $m_\phi$ is the
mass of the $\phi$ meson)
the application of hydrodynamics may be questionable and 
contributions from perturbative QCD may become important. 
In fact, at low $p_T$ domain the contribution of thermal $s\bar{s}$
for $\phi$ formation is dominant~\cite{hwa}. 
In that case, 
the description for $T_i=480$ MeV is better suited. 
From the inverse  slope (for $0.45\leq p_T$ (GeV)$\leq 2$)
of the data we get the effective temperature as
$257\pm 8$ MeV. We find that the average radial velocity 
$<v_r>\sim 0.43$ at the freeze-out surface. Therefore, using the
relation $T_{eff}=T_{true}+(1/2)m_\phi<v_r>^2$, we get the `true' 
freeze-out temperature $T_{true}\sim 162\pm 8$ MeV (error bars
correspond to the experimental error bars~\cite{STAR}). It should be
noted that the value of $T_{true}$ is close to $T_c$ predicted by
lattice QCD simulation.  The value of the hadronization temperature 
in statistical hadronization model~\cite{jrafelski} was found as
$T_c=155\pm 8$ MeV (see also~\cite{pbm}). The similar value
of hadronization temperature was obtained ~\cite{gorenstein} 
from the analysis of
experimental data on heavier hadrons.
%Furthermore,  
%the value of average radial velocity is larger at the point
%of kinetic freeze-out for pions~\cite{nuxu}.

Now we  turn to  the yield of $\phi$ mesons at mid 
rapidity. The experimental value of $\frac{dN_{\phi}}{dy}$ 
for 0-5\% central collisions is 7.7 $\pm$ 0.30~\cite{STAR}. 
It is more convenient 
to deal with the ratio, $\frac{dN_{\phi}}{dy}$/$\frac{dN}{dy}$, 
where  $dN/dy$ is the total particle multiplicity at mid-rapidity. 
This ratio has the advantage in the sense that some of the uncertainties 
({\it e.g.} volume)are removed. From 
thermal model calculations it can be shown that at the critical 
temperature $T_{c}$,

\be
\noindent \frac{\frac{dN_{\phi}}{dy}}{\frac{dN}{dy}} = 0.7 \frac{1}{g_{eff}(T_{c})}
\int^{\infty}_{M_{\phi}/T_{c}}{ \frac{x^{2} dx}{e^{x} - 1}}.
\label{eq1}
\ee

where, $m_{\phi}$=1.02 GeV, 
and $g_{eff}$ is the effective
statistical degeneracy. Eq.~\ref{eq1}  indicates that the ratio 
$\frac{dN_{\phi}}{dy}$/$\frac{dN}{dy}$ 
depends on $T_c$ and the statistical degeneracy 
where the $\phi$ mesons freeze out.
Taking $T_{c}$ and $g_{eff}$ from lattice QCD simulations~\cite{lattice} 
as 170 MeV and 18 respectively,  
we get $\frac{dN_{\phi}}{dy}$/$\frac{dN}{dy}$ 
= 4.77$\times$ $10^{-3}$ from Eq.~\ref{eq1}. 
The experimental value of $\frac{dN_{\phi}}{dy}$/$\frac{dN}{dy}$ is 
7.7$\times 10^{-3}$. This indicates that there is a clear 
over production of $\phi$ mesons.  The data can be expressed 
in terms of the calculated value of $dN_\phi/dy\mid_{eq}$ by
a simple relation: 
\be
\gamma_\phi=\frac{dN_\phi/dy|_{expt}}{dN/dy|_{eq}}
\ee
$dN_\phi/dy|_{eq}$ 
is the value of $\phi$ meson multiplicity
expected from thermal model at $T_c$ and $\gamma_\phi$ is the enhancement 
factor (a value of $\gamma_\phi=1$ ($<1$)
indicates equilibrium (under production). The experimental data 
requires the value of $\gamma_\phi$ to be $\sim 1.3$.
This indicates that there seems to be
a over saturation of strangeness production at the temperature where
$\phi$ freezes out.

\begin{figure}
\begin{center}
\includegraphics[scale=0.5]{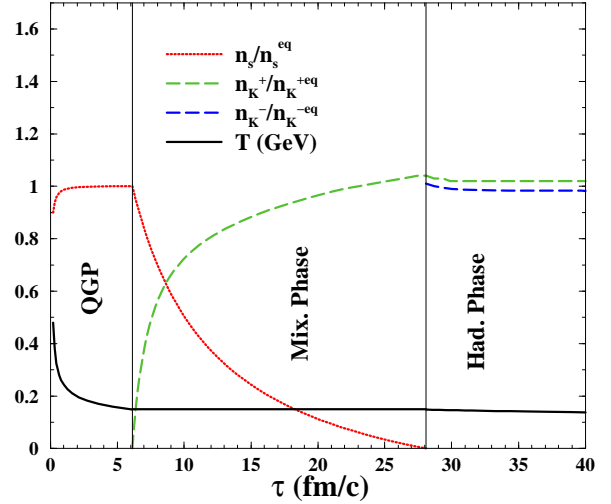}
\caption{
The time evolution of the ratio of non-equilibrium to 
equilibrium number density of strange quarks/hadrons for  
transition temperature $T_c=150$ MeV. The variation of temperature
with time is also shown.
}
\label{fig2}
\end{center}
\end{figure}

\begin{figure}
\begin{center}
\includegraphics[scale=0.5]{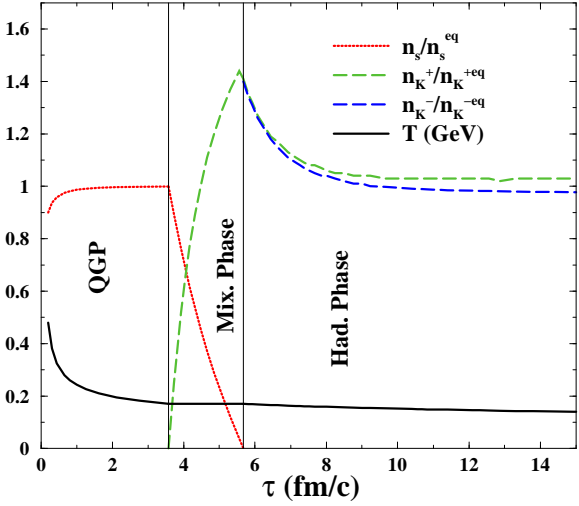}
\caption{
Same as Fig.\ref{fig2} for $T_c=170$ MeV.
}
\label{fig3}
\end{center}
\end{figure}

\begin{figure}
\begin{center}
\includegraphics[scale=0.5]{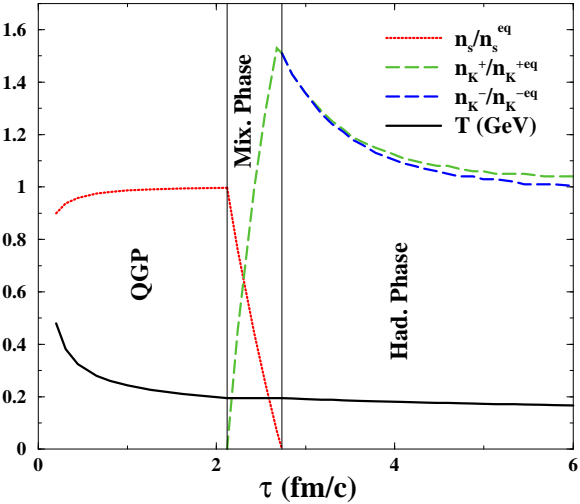}
\caption{
Same as Fig.\ref{fig2} for $T_c=195$ MeV.
}
\label{fig4}
\end{center}
\end{figure}

Now we ask, at what stage of the evolution  
of the system formed in heavy-ion collisions gives the required over production
of strangeness?  We adopt the momentum integrated Boltzmann equation
to study the evolution of the strangeness density. 
This has also been used extensively to investigate the density 
evolution in the early universe~\cite{kolb}.
The initial state is considered
as a thermalized state of gluons and light quarks with initial temperature,
$T_i\sim 480 $ MeV and thermalization time, $\tau_i\sim 0.2$ fm/c. The 
reactions considered for the  strange quark 
productions in QGP are:
$q\bar{q}$ $\rightarrow$ $s\bar{s}$ and 
$gg$ $\rightarrow$ $s\bar{s}$. For the production
of strange  hadrons the following reactions are considered  
~\cite{lileebrown}:
$\pi^{+} + \pi^{-}$  $\rightarrow$ $K^{+} + K^{-}$,
$\pi^{+} + \pi^{0}$  $\rightarrow$ $K^{+} + \bar{K^{0}}$ and
$\pi^{0} + \pi^{0}$  $\rightarrow$ $K^{+} + K^{-}$ in the 
meson-meson sector and 
$\pi^{+} + n$  $\rightarrow$ $\Lambda + K^{+}$,
$\pi^{0} + p$  $\rightarrow$ $\Lambda + K^{+}$,
$\pi^{+} + p$  $\rightarrow$ $\Sigma^{+} + K^{+}$,
$\pi^{-} + p$  $\rightarrow$ $\Sigma^{-} + K^{+}$ and
$\pi^{0} + n$  $\rightarrow$ $\Sigma^{-} + K^{+}$ for
the meson-baryon sector.

The evolution of $\bar{s}$ in QGP phase with proper time $\tau$  
is given by

\begin{equation}
\frac{dr_{\bar{s}}}{d\tau} = \frac{R_{\bar{s}}(T)}{n^{eq}_{\bar{s}}}
[ 1 - r_{s}r_{\bar{s}}] 
\label{eq2}
\end{equation} 
Similar equation can be written down for $s$ quarks. In the present
case  $r_s=r_{\bar{s}}$.
The evolution of $K^+ (u\bar{s}$ mesons 
in the mixed phase is governed by
the following equation:

\bea
\frac{dr_{K^{+}}}{d\tau}&=&\frac{R_{K^{+}}(T_c)}{n^{eq}_{K^{+}}}
\left(1 - r_{K^{+}}r_{K^{-}}\right)\nn\\
&+&\frac{R_{\Lambda}(T_c)}{n^{eq}_{K^{+}}}
\left(1-r_{K^{+}}r_{\Lambda}\right)\nn\\
&+&\frac{R_{\Sigma}(T_c)}{n^{eq}_{K^{+}}}
\left(1-r_{K^{+}}r_{\Sigma}\right)\nn\\
&+&\frac{1}{f}\frac{df}{dt}\left(\alpha\frac{r_{\bar{s}}n^{eq}_{\bar{s}}}
{n^{eq}_{K^{+}}}
- r_{K^{+}}\right)
\label{eq3}
\eea

In the above equation the last term stands for the hadronization of 
$\bar{s}$ quarks to $K^+$~\cite{kapusta,matsui}. $\alpha$ is a
parameter which indicates the fraction of $s$ ($\bar{s}$) quarks 
hadronizing to $K^-$ ($K^+$), the value of $\alpha=0.5$
if we consider  $K^+$, $\bar{K^0}$, $K^-$ and $K^0$ formation
in the mixed phase.  $r_i$ denotes
the ratio of non-equilibrium ($n_i$) to equilibrium ($n_i^{eq}$) density
of the particle $i$.
Similar equation can be written for $r_k$ in the hadronic phase. 
The evolution of other particles {\it e.g.} 
$K^-$ can be treated similarly with appropriate cross 
sections~\cite{lileebrown}
as inputs and solving the coupled set of equations. The initial value
of $s$ and $\bar{s}$ quarks are taken close to their equilibrium value. 
However, a small change in the initial values of $r_s$ 
do not change the final results
drastically. Even with lower initial values of $s$ and $\bar{s}$ 
the system reaches equilibrium very fast due to their production
in the high temperature heat bath. 
In Eq.~\ref{eq3} $f$ and $(1-f)$ are the fractions of hadrons and QGP 
respectively in the mixed phase.
$R_i(T)$ is the production rate of particle $i$ at temperature $T$.
The cooling of the heat bath is governed by the following equation~\cite{ijmpa}:
\bea
\frac{dT}{d\tau}&=&-c_s^2\frac{T}{\tau}\nn\\
&-&\frac{b(\dot{r_s}+\dot{r_{\bar{s}}})}{\alpha(a+b(r_s+r_{\bar{s}}))}
\label{eq4}
\eea
where $\alpha=(1+c_s^2)/c_s^2$, $a=8\pi^2/(45c_s^2)$ and 
$b=7\pi^2n_F/(120c_s^2$),  $c_s^2$ is the velocity of sound, $n_F$
is the number of flavours and
$\dot{r_s}=dr_s/d\tau$. 
It may be noted that the Bjorken's cooling law can be recovered
from Eq.~\ref{eq4} when the second term vanishes. The value of
net baryon number at the central rapidity has been assumed to be zero.

The time evolutions of the
ratios of non-equilibrium to equilibrium number density ($r_{i}$) 
for $T_c=$ 150, 170 and 195 MeV 
are shown in Fig.~\ref{fig2},\ref{fig3} and \ref{fig4}
respectively. It should be mentioned here that the corresponding
$g_{eff}$ for the above values of temperature are
taken from lattice QCD~\cite{lattice}. 
We observe a clear over saturation in number density of kaons
at the end of the  mixed phase (Fig.~\ref{fig3}\& \ref{fig4}). 
For $T_c=170 - 195$ MeV the value of 
$r_{K}$ is 1.4 - 1.5  
at the end of the mixed phase, before it reaches the equilibrium value
of 1 in the hadronic phase. It may be noted that there is a 
small difference between $r_{K^+}$ and $r_{K^-}$ in the
hadronic phase because of their different production cross sections.   
For the value of $T_{c}$ = 150 MeV, we observe that $r_{K^+}$ $\sim$ 1,
for $\alpha=0.5$. For smaller values of $\alpha$, $r_{K^+}$ will
be even smaller. 

Considering that we need a  $\gamma_{\phi}> 1$ to explain the experimental
data (as explained before) our calculations  set a lower bound
on the value of $T_{c}$, it must be $>$ 160 MeV.
For a hadronic initial state  with temperature of 250 MeV 
the value of  $r_{K}$ remains $\leq$ 1.
This indicates that the
enhanced production of $\phi$ mesons in heavy-ion collisions 
at $\sqrt{s_{NN}}$ = 200 GeV can be explained if the
system passes through a mixed phase during it's evolution with a
$T_{c}$ $>$ 160 MeV. At this value of $T_c$, the
enhancement factor, $\gamma_\phi\sim
r_{K^+}\times r_{K^-}\sim \gamma_{K}^2 \sim 1.24$.
Considering the fact that the experimental data indicate
$\gamma_\phi\sim 1.4\pm 0.1$ we argue that $T_c > 160$ MeV.

The life time of the mixed phase crucially depends on the effective
statistical degeneracy of QGP and hadronic phases. $g_{eff}\sim 9, 26$ and
31 for $T=150, 170$ and 195 MeV respectively. Larger difference  in
the statistical degeneracy gives rise to larger discontinuity in the
entropy density which consequently generates more latent heat. In the 
mixed phase the cooling due to expansion is compensated by the
liberation of latent heat. Hence if larger latent is generated 
then the system will spend more time in the mixed phase {\it i.e.}
the life time of the mixed phase will be larger. This is clearly 
reflected in the results shown in Fig.~\ref{fig2}-\ref{fig4}.

\begin{figure}
\begin{center}
\includegraphics[scale=0.5]{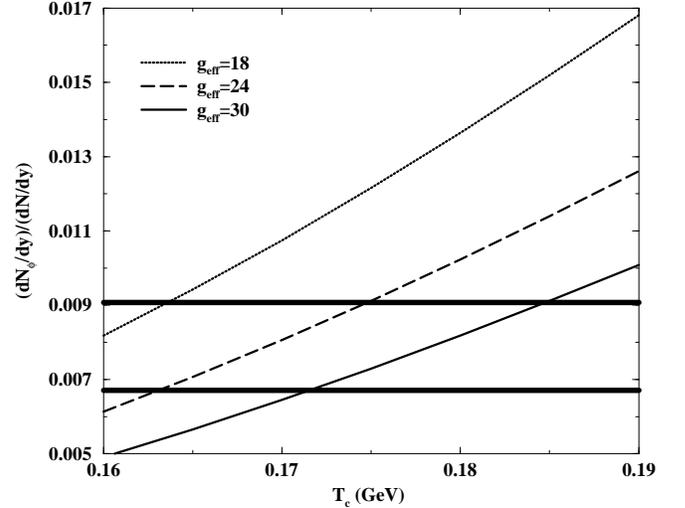}
\caption{
The variation of $(dN_\phi/dy)/(dN/dy)$ with $T_c$ for three values
of $g_{eff}$.
The thick solid horizontal lines represents experimental 
value of $(dN_\phi/dy)/(dN/dy)$ with associated error bars.
}
\label{fig5}
\end{center}
\end{figure}

We characterize the mixed phase in terms of effective
statistical degeneracy $g_{eff}$. This is justified
because, (i) $\phi$ mesons freeze out at  $T_c$ and (ii)
its production can only be explained by the existence of
mixed phase.  For fixed $g_{eff}$
we calculate $\frac{dN_{\phi}}{dy}$/$\frac{dN}{dy}$ from Eq.~\ref{eq1}
This ratio is then multiplied by
$\gamma_\phi$ and compared with the corresponding value from experimental
data (lies within the two thick horizontal lines in Fig.~\ref{fig5}). 
The results for $g_{eff}$ =
18, 24, and 30 are shown in Fig.~\ref{fig5}. We find that  
$g_{eff}\sim 18$ may be considered as a lower bound.

%\section{Summary}
In summary, we have studied the production of $\phi$ mesons in Au+Au collisions
at $\sqrt{s_{NN}}$ = 200 GeV. The transverse momentum spectra is well explained
by  hydrodynamical model with the equation of state taken from lattice QCD.
It is observed that
$\phi$ mesons freezes out around the transition temperature 
predicted by lattice QCD.
The over production of $\phi$ meson at mid-rapidity can only be explained 
if QGP is formed in the initial state and passes 
through a co-existence  phase of QGP and hadrons, {\it i.e.}  
over saturation in strangeness indicates the formation of co-existing
phase of QGP and hadrons (indications of which has already been pointed out at
SPS energies~\cite{us,gazdzicki}). 
The analysis of the 
experimental data reveals a lower bound on the  transition temperature
and  the effective statistical degeneracy. 
The extracted value of $T_c\sim162\pm 8$
and the lower bound on $g_{eff}\sim$ 18.
The study of $K/\pi$ in the similar framework is under progress~\cite{jajati}.

\end{document}